# Tackling the 6/49 Lottery and Debunking Common Myths with Probabilistic Methods and Combinatorial Designs

Ralph Stömmer *

* Private Researcher

e-mail: ralph.stoemmer@trajectorix.de

(Working paper, version 3, May 29, 2026)

**Abstract**

At the end, the house always wins! This simple truth holds for all public games of chance. Nevertheless, since lotteries have existed, people have tried everything to give luck a helping hand. This article compares objective scientific approaches to tackle the 6/49 lottery: probabilistic methods and combinatorial designs. The mathematical models developed herein can be modified and applied to other lotteries. The newly constructed (49, 6, 5) covering design is introduced, which meets the Schönheim bound. For lottery designs and for covering designs, a benchmark based on probabilistic methods is presented. It is demonstrated that common attempts to outwit the odds correspond to limitations of numbers to subsets, which disproportionately reduce the chances of winning.

## 1. Introduction

There are many lotteries around the world. A common lottery scheme is (49, 6, 6, t), known as 6/49 lottery, where 6 numbers are randomly drawn by the lottery agency from 49 numbers without replacement. Beforehand, people buy tickets and choose 6 numbers per ticket. People achieve a t-hit if the 6 numbers drawn intersect the 6 numbers on their ticket in t common elements. The prize depends on the country, the lottery agency, and the rules of the game. In general, the prize scales with t. For some lotteries the prize starts at t = 3, sometimes sufficient to cover the ticket costs, up to t = 6 for the jackpot.

There are $\binom{49}{6}$ = 13,983,816 possible combinations to choose 6 from 49 numbers. From that large quantity, the jackpot is just a single combination. The calculation of its probability P(6) is intuitively clear: P(6) = 1 / 13,983,816 = 0.0000072%. Despite the extremely low value, a lot of people try their luck, on occasion on a weekly basis. Assuming that 13 million players, each with a single ticket, would participate at a nationwide draw, one should expect 13 million x P(6) ≈ 1 winner on average. If the lucky one tells news media that he produced the winning combination from birthdays, questionable attributions can backfire.

There is a more sophisticated evaluation of odds. Let us start with a warm-up exercise on complementary probabilities, which are crucial in determining probabilities with not just one, but with many lottery tickets. It turns out that probabilities for single tickets are well known, but work published on many tickets is rare. Let P(6) again denote the probability to achieve a 6-hit, the jackpot, with a single ticket. The complementary probability, to achieve no 6-hit, calculates to [1 – P(6)] = 99.99999%. Ending up empty-handed with a single ticket is nothing new. Let´s consider the unrealistic case that we were able to submit v = 5 million tickets, filled out at random. To achieve no 6-hit at all with 5 million tickets calculates to $[1 - P(6)]^{5\text{million}} \approx 70\%$. Again, we take the complementary value to obtain the probability to win the jackpot with v = 5 million tickets: P(jackpot, v = 5 million) = $1 - [1 - P(6)]^{5\text{million}} \approx 30\%$. That is a fairly good chance of winning, if bulk submission of tickets was still allowed and technically feasible. However, such loopholes were closed in the past, after syndicates were successfully tackling lotteries with ticket quantities and logistic systems beyond the capabilities of individuals [1]. At the end, the house always wins.

What is left for the player? The choice of the numbers on each ticket, and the number of tickets, up to a maximum quantity limited by the lottery agency. However, there is still some room for maneuver. The next chapter reveals the scientific world hidden in between these two options.

## 2. Theory and application

### *2.1. Tackling the 6/49 lottery with probabilistic methods*

Although the basics of probabilistic methods are widely known, some insights have not been widely disseminated. The probability P(t) to achieve a t-hit is given by the fraction $P(t) = M_t/N$, the classical definition of probability. N denotes the total number of combinations, $N = \binom{n}{p}$, when p numbers are randomly drawn from a total of n numbers. $M_t$ denotes the number of all possible combinations with a t-hit. It is given by

$$M_t = \binom{p}{t}\binom{n-p}{p-t}\,. \tag{1}$$

The total number N of combinations can alternatively be obtained with the sum over $M_t$:

$$\sum_{t=0}^{p} M_t = N\,. \tag{2}$$

Table 1 lists all possible combinations $M_t$ with a t-hit with their corresponding probabilities P(t) for the 6/49 lottery, with n = 49, p = 6, and index t = 0, …, 6. The probabilities match with the experience of players. In the majority of cases, 0-hits up to 2-hits are achieved. Sometimes 3-hits occur, and 4-hits are rare. A lot of luck is required to achieve a 5-hit, and it is more likely to be struck by lightning than to achieve a 6-hit and win the jackpot.

| t | 0 | 1 | 2 | 3 | 4 | 5 | 6 |
|---|---|---|---|---|---|---|---|
| Mt | 6,096,454 | 5,775,588 | 1,851,150 | 246,820 | 13,545 | 258 | 1 |
| P(t) | 43.596% | 41.302% | 13.238% | 1.765% | 0.097% | 0.002% | 0.0000072% |

Table 1: All combinations $M_t$ with a t-hit and probabilities P(t) for the 6/49 lottery.

In the following sections, we elaborate on probabilities to achieve 5-hits and the 6-hit with many tickets. If required, the theory can be extended including 4-hits or even 3-hits. By changing the parameters n and p, the theory can be adapted to other lotteries.

### *2.1.1. The probabilistic method with unique tickets avoiding doubles*

Let us now discuss the strategy in which we pick v tickets at random, but making sure that each tickets is unique (no doubles, all tickets have to be pairwise different). $M_t$ denotes the number of all possible combinations with a t-hit, as provided in equation 1. It is obvious that by filling out more than just one ticket, one might achieve more than just one t-hit listed in table 1. A more generalized theory needs to be applied, tailored to the quantity v of tickets. Let $P(m_t, v)$ denote the probability to achieve exactly $m_t$ t-hits, when v tickets are chosen at random with no doubles. It is given by the hypergeometric distribution

$$P(m_t, v) = \frac{\binom{M_t}{m_t}\binom{N-M_t}{v-m_t}}{\binom{N}{v}} \ . \tag{3}$$

In order to explore the structure in equation 3, we may check familiar cases. For instance, $P(m_t = 1, v = 1)$ is the probability to achieve exactly one t-hit with a single ticket:

$$P(m_t = 1, v = 1) = \frac{\binom{M_t}{1}\binom{N-M_t}{0}}{\binom{N}{1}} = \frac{M_t}{N} \ . \tag{4}$$

The fraction $M_t/N$ on the right hand side in equation 4 is the probability P(t) already introduced in the previous section. If we want to know the probability to achieve the 6-hit and several 5-hits with a certain quantity v of tickets, the multivariate hypergeometric distribution is required. $P(m_5, m_6, v)$ denotes the probability to achieve $m_5$ 5-hits and $m_6$ 6-hits with v tickets:

$$P(m_5, m_6, v) = \frac{\binom{M_5}{m_5}\binom{M_6}{m_6}\binom{N-M_5-M_6}{v-m_5-m_6}}{\binom{N}{v}} \ . \tag{5}$$

$M_5 = 258$, $m_5$ can take values in the range 0 … 258, depending on how many tickets v are filled out. Since $M_6 = 1$, $m_6$ can only take the values 0 or 1. Equation 5 serves to calculate the complementary probability for achieving a minimum number of t-hits. At first, we ask for the probability that we achieve no 5-hit and no 6-hit at all with v tickets. With equation 5 and $m_5 = 0$, $m_6 = 0$, we get

$$P(m_5 = 0, m_6 = 0, v) = \frac{\binom{M_5}{0}\binom{M_6}{0}\binom{N-M_5-M_6}{v-0-0}}{\binom{N}{v}} = \frac{\binom{N-M_5-M_6}{v}}{\binom{N}{v}} \ . \tag{6}$$

The fraction on the right hand side of equation 6 does consist of two binomial coefficients, which for larger values are challenging to compute. There are two useful approximations of equation 6, for small v and for small $(M_5 + M_6)$.

Condition $v < N - (M_5 + M_6)$:

$$P(m_5 = 0, m_6 = 0, v) = \frac{\binom{N-M_5-M_6}{v}}{\binom{N}{v}} \approx \left(1 - \frac{M_5+M_6}{N}\right)^v . \quad (7)$$

Condition $(M_5 + M_6) < (N - v)$:

$$P(m_5 = 0, m_6 = 0, v) = \frac{\binom{N-M_5-M_6}{v}}{\binom{N}{v}} \approx \left(1 - \frac{v}{N}\right)^{M_5+M_6} . \quad (8)$$

The approximation in equation 7 is identical to the expression gained for filling out tickets at random, allowing for doubles (see equation 11 in the following chapter). As long as the condition $v < N - (M_5 + M_6)$ holds, the number of tickets is sufficiently small that it doesn´t make a difference if tickets are unique without doubles or not.

For the conditions of the 6/49 lottery and larger quantities v of tickets, equation 8 is better suited, since $(M_5 + M_6) = 259$, which is small compared to $(N - v)$. The complementary probability to achieve at least one 5-hit with v tickets is given by

$$P(m_5 \geq 1, v) = 1 - P(m_5 = 0, m_6 = 0, v) \approx 1 - \left(1 - \frac{v}{N}\right)^{M_5+M_6} . \quad (9)$$

To be more specific about the argument in brackets provided in $P(m_5 \geq 1, v)$: achieving at least one 5-hit includes 1 or 2 or 3 … or finally all $M_5 = 258$ combinations with 5-hits or the 6-hit (with $M_6 = 1$ there is only one 6-hit).

One should internalize the difference between “exactly” one 5-hit, and “at least” one 5-hit. The probability to achieve exactly one 5-hit with v = 60,000 tickets is calculated with equation 3, which yields $P(m_5 = 1, v = 60{,}000) \approx 37\%$. In contrast to that, the probability to achieve at least one 5-hit with 60,000 tickets is calculated with equation 9, which yields $P(m_5 \geq 1, v = 60{,}000) \approx 67\%$. The value is higher, because it is not limited to just one 5-hit. It does include other possible events such as two 5-hits and no 6-hit up to 258 5-hits and the 6-hit. Equation 9 is applied to determine the number of tickets required to achieve specific probabilities for at least one 5-hit. Some values are provided in table 2.

| (49, 6, 6, 5) lottery scheme | | | | | | | | |
|---|---|---|---|---|---|---|---|---|
| Number of tickets v | 1 | 10 | 100 | 1,000 | 10,000 | 100,000 | 300,000 | 500,000 |
| Probability for at least one 5-hit $P(m5 \geq 1, v)$ | 0.002% | 0.019% | 0.185% | 1.835% | 16.913% | 84.414% | 99.636% | 99.992% |

Table 2: Probabilities to achieve at least one 5-hit related with the number of tickets.

The expression for the 6-hit is derived from equation 9. With $M_5 = 0$ and $M_6 = 1$ we get

$$P(m_6 = 1, v) = 1 - P(m_6 = 0, v) = 1 - \left(1 - \frac{v}{N}\right)^1 = \frac{v}{N} \ . \qquad (10)$$

With unique tickets (no doubles, all blocks have to be pairwise different), the probability to win the jackpot scales linearly with the number of tickets v. The identical result can be obtained with equation 3. The only chance to secure the jackpot with no risk at all is “buying the pot”, which stands for submitting all possible combinations, v = N.

The methods above can be extended in a straightforward manner including 3-hits and 4-hits. One just needs to extend equations 5 to 9 with $M_3$ and $M_4$. By changing the parameters n and p, the theory can be adapted to other lotteries.

*2.1.2. The probabilistic method with tickets allowing for doubles*

Let us now discuss the strategy in which we pick v tickets at random, without making sure that each ticket is unique, allowing for doubles. In other words: the player doesn´t care about the numbers on previously filled out tickets. It simplifies the reasoning to switch the parameters from $M_t$ and N to the probabilities P(t).

The probability to achieve a 5-hit or the 6-hit with a single ticket is P(5) + P(6). The complementary probability to achieve no 5-hit and no 6-hit with a single ticket calculates to 1 – P(5) – P(6), and the probability to achieve no 5-hit and no 6-hit with v tickets is

$$P(m_5 = 0, m_6 = 0, v) = [1 - P(5) - P(6)]^v \; . \qquad (11)$$

Equation 11 is identical to equation 7. The probability $P(m_5 \geq 1, v)$ to achieve at least one 5-hit with v tickets calculates to

$$P(m_5 \geq 1, v) = 1 - [1 - P(5) - P(6)]^v \; . \qquad (12)$$

For the 6/49 lottery, the results obtained with equation 12 differ only marginally from that obtained with equation 9 in table 2. For targeting at least one 5-hit, tickets can be filled out allowing for doubles or not. The expression for achieving the 6-hit is

$$P(m_6 = 1, v) = 1 - [1 - P(6)]^v \; . \qquad (13)$$

Assuming that it was possible to submit 5 million tickets allowing for doubles, we obtain the result briefly discussed in the introduction:

$$P(m_6 = 1, v = 5{,}000{,}000) = 1 - \left[1 - \frac{1}{13{,}983{,}816}\right]^{5{,}000{,}000} = 30.062\% \; .$$

However, for increasing the probability to achieve the 6-hit it is smarter to fill out tickets with no doubles. Equation 10 yields

$$P(m_6 = 1, v = 5{,}000{,}000) = \frac{5{,}000{,}000}{13{,}983{,}816} = 35.756\% \; .$$

## *2.2. Tackling the 6/49 lottery with combinatorial designs*

### *2.2.1. The lottery design with benchmark*

An (n, k, p, t) lottery scheme consists of an n-set of elements V = {1, 2, 3, …, n}, a collection of k-element subsets of V (called k-subsets or blocks), and a p-element subset of V (called p-subset). The parameters n, k, p, t must satisfy the relationship $t \leq \{p, k\} \leq n$. The k-subsets, or blocks, are designated as a lottery design LD(n, k, p, t) when any p-subset of V intersects at least one of these blocks in at least t elements. The smallest quantity of blocks of the lottery design LD(n, k, p, t) is denoted as lottery number L(n, k, p, t) [2]. In lotteries, the blocks correspond to the tickets.

Large lottery designs to address the 6/49 lottery haven´t been around for long, because they are challenging to compute. Before explicit designs came up, the lower and the upper bounds for the lottery number L(49, 6, 6, 5) were determined to at least narrow down the challenge. In this context, the work of W. R. Gründlingh is significant, because it provides a survey of achievements up to 2004 [3]. L(49, 6, 6, 5) = 62,151 is provided as the lower bound, which is the number of tickets to be filled out for achieving at least one 5-hit. This lower bound has not been reached so far. The current record for the lottery design LD(49, 6, 6, 5) was uploaded in April 2026 into the online database “Covering Repository” [4]. It does contain 142,351 blocks, which is the number of tickets to be submitted for achieving at least one 5-hit with absolute certainty.

The lottery design should be benchmarked with equations 9 and 10, where the same number of tickets was filled out at random, avoiding doubles. With 142,351 tickets the probability to achieve at least one 5 hit calculates to

$$\mathrm{P}(\mathrm{m}_5 \geq 1,\ \mathrm{v} = 142{,}351) = 1 - \left(1 - \frac{142{,}351}{13{,}983{,}816}\right)^{258+1} = 92.935\% \,.$$

With 142,351 tickets, the residual risk of ~7% to miss the 5-hit might look acceptable for bold players, if they dispensed with the lottery design and rather try the probabilistic method. For the 6-hit, we obtain from equation 10:

$$\mathrm{P}(\mathrm{m}_6 = 1, \mathrm{v} = 142{,}351) = \frac{142{,}351}{13{,}983{,}816} = 1.018\% \,.$$

When the lower limit of 62,151 blocks might be reached for the LD(49, 6, 6, 5) lottery design at some point in the future, things look different. The probability to achieve the 6-hit will decrease, for the lottery design as well as for the probabilistic method, because it just scales with the number v of submitted tickets. The probability to achieve at least one 5-hit would be

$$P(m_5 \geq 1,\ v = 62{,}151) = 1 - \left(1 - \frac{62{,}151}{13{,}983{,}816}\right)^{258+1} = 68.453\% \ .$$

In this case, applying the lottery design was a clear advantage, if the player was able to submit 62,151 tickets, and if the profit was higher than the costs for submission. The probabilistic method just guarantees a probability of ~68%, whereas the lottery design guarantees an absolute certainty to achieve at least one 5-hit.

### *2.2.2. The covering design with benchmark*

A (n, k, t) covering design is a collection of k-element subsets (called k-subsets or blocks) of an n-set of elements V = {1, 2, 3, …, n} such that any t-element subset (called t-subset) is contained in at least one block. The parameters n, k, t must satisfy the relationship $t \leq k \leq n$. The smallest quantity of blocks of the (n, k, t) covering design is denoted as covering number C(n, k, t) [5, 6]. In lotteries, the blocks correspond to the tickets.

Covering designs and lottery designs are related to each other [7]. Adding the trivial case that all k-subsets of an n-set, with gives $N = \binom{n}{k}$ combinations in total, do provide the biggest covering design, L(n, k, p, t) and C(n, k, t) fulfill the inequalities

$$L(n, k, p, t) \leq C(n, k, t) \leq \binom{n}{k} \ . \tag{14}$$

By construction covering designs are larger than lottery designs. In contrast to lottery designs, which just guarantee at least one 5-hit, covering designs are complete insofar as they guarantee at least all six 5-hits of a random draw of 6 numbers in the 6/49 lottery.

Covering designs can be obtained from "The La Jolla Covering Repository", a comprehensive online database for combinatorial structures [8]. The (49, 6, 5) covering design was constructed and added to the database in February 2026 [9]. It is also made available in the "Covering Repository" [4]. It is built on the groundwork laid by R. H. F. Denniston and A. Jurcovich [10, 11]. C(49, 6, 5) = 325,205 does meet the Schönheim bound, which denotes the lower limit of covering designs [12]. That means: in order to cover all six 5-hits in the 6/49 lottery, which includes the case that at least one 5-hit is achieved, one has to submit the 325,205 tickets as provided in the (49, 6, 5) covering design. There is no smaller, no more efficient covering design.

As accomplished in the previous chapter with the lottery design, the covering design is benchmarked with a random test with equations 9 and 10. If 325,205 tickets were filled out at random avoiding doubles, the probability to achieve at least one 5 hit calculates to

$$\mathrm{P}(\mathrm{m}_5 \geq 1,\ \mathrm{v} = 325{,}205) = 1 - \left(1 - \frac{325{,}205}{13{,}983{,}816}\right)^{258+1} = 99.774\% \ .$$

The probability for the 6 hit just scales with the number of tickets submitted, for the probabilistic method and for the covering design alike. From equation 10 we get:

$$\mathrm{P}(\mathrm{m}_6 = 1, \mathrm{v} = 325{,}205) = \frac{325{,}205}{13{,}983{,}816} = 2.326\% \ .$$

With 325,205 tickets, the residual risk of ~0.2% to miss any 5-hit is negligible – if we dispense with the covering design and go for the probabilistic method. However, the covering design does guarantee at least all six 5-hits in the 6/49 lottery. When choosing random tickets avoiding doubles, such an event does happen less often. To cover at least all six 5-hits calculates to $\mathrm{P}(\mathrm{m}_5 \geq 6, \mathrm{v} = 325{,}205) \approx 56\%$ only. If the prize for 5-hits was high enough and bulk submission was still possible, one should submit tickets filled out according to the (49, 6, 5) covering design.

Lottery agencies have prohibited bulk submissions, and the prizes for 5-hits ensure that the expected profit is not worth the efforts. Although the (49, 6, 5) covering design is finally available, its relevance is purely academic – due to its efficiency a mathematical beauty, so to speak. It might be applied as a building block for other designs of interest.

### 2.3. Debunking myths

On occasion, jackpot winners do attribute their luck to birthday numbers or to any number patterns on tickets. People who have never won anything at the lottery do that, too. The mathematics in the previous chapters allows to debunk such myths. Any restriction of numbers to a few lucky numbers, to even or to odd numbers, to number patterns such as diagonals across the 7 x 7 square field of 49 numbers, or to the left, to the right, to the upper or to the lower half of the square field can mathematically be treated as a n*-subset out of 49 numbers. Whatever its position in the square field, no number stands out from all the other numbers.

In order to quantify the restriction to n* numbers, we ask: what is the probability, that at least 5 of the 6 winning numbers fall among the n* picks chosen from n = 49 numbers? The logic behind is equal to asking for the probability that a random draw of p = 6 balls from an urn of n = 49 balls, which contains n* white balls and (n - n*) black balls, does include t = 5 or t = 6 white balls. The answer is given by the hypergeometric distribution

$$P(t, n^*) = \frac{\binom{n^*}{t}\binom{n-n^*}{p-t}}{\binom{n}{p}} \quad . \tag{15}$$

Equation 15 has the same structure as equation 3, but one should internalize that the perspective has turned to the p = 6 winning numbers. With t = 6, p = 6, and n = 49 equation 15 yields

$$P(t=6,n^*)=\frac{\binom{n^*}{6}\binom{49-n^*}{0}}{\binom{49}{6}}=\frac{\binom{n^*}{6}}{\binom{49}{6}}\ . \tag{16}$$

With t = 5, p = 6, and n = 49 equation 15 yields

$$P(t=5,n^*)=\frac{\binom{n^*}{5}\binom{49-n^*}{1}}{\binom{49}{6}}=\frac{\binom{n^*}{5}(49-n^*)}{\binom{49}{6}}\ . \tag{17}$$

That at least 5 winning numbers, which includes both cases t = 5 and t = 6, fall among the n* picks is the sum of equations 16 and 17:

$$P(t\geq 5,n^*)=\frac{\binom{n^*}{6}}{\binom{49}{6}}+\frac{\binom{n^*}{5}(49-n^*)}{\binom{49}{6}}\ . \tag{18}$$

Example A: If one applies n* = 10 favorite numbers, for instance mixed from various birthdays, the probability that at least 5 winning numbers fall among the 10-subset calculates to

$$P(t\geq 5,n^*=10)=\frac{\binom{10}{6}}{\binom{49}{6}}+\frac{\binom{10}{5}(49-10)}{\binom{49}{6}}=\ 0.072\%\ .$$

The probability is extremely low. It takes effect only if all combinations of the 10-subset, which calculates to $\binom{10}{6}$ = 210 tickets, are filled out. However, the tickets would be better spent on all 49 numbers. Equation 9 yields

$$P(m_5\geq 1,v=210)=1-\left(1-\frac{210}{13{,}983{,}816}\right)^{258+1}=0.388\%\ .$$

Example B: If one focuses onto combinations in one half of the square field of 49 numbers, which amounts to n* = 25, the probability that at least 5 winning numbers fall among the 25-subset calculates to

$$P(t\geq 5,n^*=25)=\frac{\binom{25}{6}}{\binom{49}{6}}+\frac{\binom{25}{5}(49-25)}{\binom{49}{6}}=10.385\%\ .$$

One could naively assume a chance of ~50%, that a fair distribution of luck would scale with the quantity of numbers covered. As so often, when faced with combinatorics and probabilities, intuition is wrong. For P(t ≥ 5, n* = 25) = 10.385% to take effect, all combinations of the 25-subset have to be filled out, which calculates to $\binom{25}{6}$ = 177,100 tickets. Again, the tickets would be better spent on all 49 numbers. Equation 9 yields

$$P(m_5 \geq 1, v = 177{,}100) = 1 - \left(1 - \frac{177{,}100}{13{,}983{,}816}\right)^{258+1} = 96.316\% \ .$$

If we applied a (10, 6, 5) covering design with C(10, 6, 5) = 50 in example A, and a (25, 6, 5) covering design with C(25, 6, 5) = 9,321 in example B, calculations show that it was still better to follow the probabilistic method and just spend the 50 tickets in example A and the 9,321 tickets in example B on all 49 numbers.

The key lesson: Limiting lottery numbers to preferred subsets disproportionately reduces the probabilities to achieve at least one 5-hit. It can even compensate the positive effect of covering designs. Focusing onto specific subsets of numbers, or onto any number patterns, just reflects human errors, for which behavioral economists once coined the expressions “control illusion” and „simulation error“. People boost their self-efficacy believing they would control the outcome of random events, and arranging numbers gives the halo of expertise where none exists.

## 3. Conclusion

This work combines probabilistic methods to tackle the 6/49 lottery with lottery designs and with covering designs. As expected, none of these methods works miracles. The probabilistic method can be applied as a benchmark for combinatorial designs. If lottery agencies hadn´t closed the loopholes for bulk submissions, and prizes for 5-hits were high enough to justify the efforts, the LD(49, 6, 6, 5) lottery design should further be improved, because L(49, 6, 6, 5) = 62,151 has not been reached yet.

The newly constructed (49, 6, 5) covering design is introduced, which was not publicly available before. However, C(49, 6, 5) = 325,205 blocks could as well be filled out at random for achieving at least one 5-hit with negligible risk. Nevertheless, the covering design guarantees at least all six 5-hits of a draw of 6 numbers. Even if the prize for a single 5-hit was low, the winnings would be paid out at least six times. As mentioned earlier, bulk submissions are not possible any more, which makes the covering design just a mathematical curiosity. Due to its size, its relevance is purely scientific, because it meets the Schönheim bound.

The only way to win the jackpot with no risk at all is “buying the pot”. The probability for the 6-hit scales with v/N, for probabilistic methods and for combinatorial designs alike. Any other promises belong in the realm of phantasy.

Common myths referring to lucky numbers or to any other preferred subsets of numbers are debunked. Such methods can be attributed to the illusion to be able to improve the outcome of random events. Any picking of numbers, be it birthday numbers, be it number patterns or input from astrology, can mathematically be treated as a preferred n*-subset of the total of n numbers. Calculations demonstrate that limiting numbers to preferences disproportionately reduces the chances of winning.

## Disclosure

I have no conflicts of interest to disclose.